\documentclass[reprint,superscriptaddress,amsmath,amssymb,aps,longbibliography]{revtex4-1}
\usepackage{graphicx}
\pdfoutput=1
\usepackage{epstopdf}
\usepackage{color}
\usepackage{dcolumn}
\usepackage{bm}
\usepackage[colorlinks=true,allcolors=blue]{hyperref}
\usepackage{upgreek} 
\usepackage{wasysym}
\usepackage{nicefrac}
\usepackage{siunitx}
\usepackage{booktabs}
\usepackage[normalem]{ulem} 
\usepackage{threeparttable} 
\usepackage{tabularx}
\usepackage{hhline}
\usepackage{caption}
\usepackage{subcaption}
\usepackage{float}
\floatstyle{plaintop}
\restylefloat{table}
\usepackage{siunitx}
\usepackage{lineno}

\newcommand{\Lagr}{\mathcal{L}}

\begin{document}

\title{First direct detection constraint on mirror dark matter kinetic mixing \protect\\ using LUX 2013 data}

\author{D.S.~Akerib} \affiliation{SLAC National Accelerator Laboratory, 2575 Sand Hill Road, Menlo Park, CA 94205, USA} \affiliation{Kavli Institute for Particle Astrophysics and Cosmology, Stanford University, 452 Lomita Mall, Stanford, CA 94309, USA} 
\author{S.~Alsum} \affiliation{University of Wisconsin-Madison, Department of Physics, 1150 University Ave., Madison, WI 53706, USA}  
\author{H.M.~Ara\'{u}jo} \affiliation{Imperial College London, High Energy Physics, Blackett Laboratory, London SW7 2BZ, United Kingdom}  
\author{X.~Bai} \affiliation{South Dakota School of Mines and Technology, 501 East St Joseph St., Rapid City, SD 57701, USA}  
\author{J.~Balajthy} \affiliation{University of California Davis, Department of Physics, One Shields Ave., Davis, CA 95616, USA}  
\author{A.~Baxter} \affiliation{University of Liverpool, Department of Physics, Liverpool L69 7ZE, UK}  
\author{E.P.~Bernard} \affiliation{University of California Berkeley, Department of Physics, Berkeley, CA 94720, USA}  
\author{A.~Bernstein} \affiliation{Lawrence Livermore National Laboratory, 7000 East Ave., Livermore, CA 94551, USA}  
\author{T.P.~Biesiadzinski} \affiliation{SLAC National Accelerator Laboratory, 2575 Sand Hill Road, Menlo Park, CA 94205, USA} \affiliation{Kavli Institute for Particle Astrophysics and Cosmology, Stanford University, 452 Lomita Mall, Stanford, CA 94309, USA} 
\author{E.M.~Boulton} \affiliation{University of California Berkeley, Department of Physics, Berkeley, CA 94720, USA} \affiliation{Lawrence Berkeley National Laboratory, 1 Cyclotron Rd., Berkeley, CA 94720, USA} \affiliation{Yale University, Department of Physics, 217 Prospect St., New Haven, CT 06511, USA}
\author{B.~Boxer} \affiliation{University of Liverpool, Department of Physics, Liverpool L69 7ZE, UK}  
\author{P.~Br\'as} \affiliation{LIP-Coimbra, Department of Physics, University of Coimbra, Rua Larga, 3004-516 Coimbra, Portugal}  
\author{S.~Burdin} \affiliation{University of Liverpool, Department of Physics, Liverpool L69 7ZE, UK}  
\author{D.~Byram} \affiliation{University of South Dakota, Department of Physics, 414E Clark St., Vermillion, SD 57069, USA} \affiliation{South Dakota Science and Technology Authority, Sanford Underground Research Facility, Lead, SD 57754, USA} 
\author{M.C.~Carmona-Benitez} \affiliation{Pennsylvania State University, Department of Physics, 104 Davey Lab, University Park, PA  16802-6300, USA}  
\author{C.~Chan} \affiliation{Brown University, Department of Physics, 182 Hope St., Providence, RI 02912, USA}  
\author{J.E.~Cutter} \affiliation{University of California Davis, Department of Physics, One Shields Ave., Davis, CA 95616, USA}  
\author{L.~de\,Viveiros}  \affiliation{Pennsylvania State University, Department of Physics, 104 Davey Lab, University Park, PA  16802-6300, USA}  
\author{E.~Druszkiewicz} \affiliation{University of Rochester, Department of Physics and Astronomy, Rochester, NY 14627, USA}  
\author{A.~Fan} \affiliation{SLAC National Accelerator Laboratory, 2575 Sand Hill Road, Menlo Park, CA 94205, USA} \affiliation{Kavli Institute for Particle Astrophysics and Cosmology, Stanford University, 452 Lomita Mall, Stanford, CA 94309, USA} 
\author{S.~Fiorucci} \affiliation{Lawrence Berkeley National Laboratory, 1 Cyclotron Rd., Berkeley, CA 94720, USA} \affiliation{Brown University, Department of Physics, 182 Hope St., Providence, RI 02912, USA} 
\author{R.J.~Gaitskell} \affiliation{Brown University, Department of Physics, 182 Hope St., Providence, RI 02912, USA}  
\author{C.~Ghag} \affiliation{Department of Physics and Astronomy, University College London, Gower Street, London WC1E 6BT, United Kingdom}  
\author{M.G.D.~Gilchriese} \affiliation{Lawrence Berkeley National Laboratory, 1 Cyclotron Rd., Berkeley, CA 94720, USA}  
\author{C.~Gwilliam} \affiliation{University of Liverpool, Department of Physics, Liverpool L69 7ZE, UK}  
\author{C.R.~Hall} \affiliation{University of Maryland, Department of Physics, College Park, MD 20742, USA}  
\author{S.J.~Haselschwardt} \affiliation{University of California Santa Barbara, Department of Physics, Santa Barbara, CA 93106, USA}  
\author{S.A.~Hertel} \affiliation{University of Massachusetts, Amherst Center for Fundamental Interactions and Department of Physics, Amherst, MA 01003-9337 USA} \affiliation{Lawrence Berkeley National Laboratory, 1 Cyclotron Rd., Berkeley, CA 94720, USA} 
\author{D.P.~Hogan} \affiliation{University of California Berkeley, Department of Physics, Berkeley, CA 94720, USA}  
\author{M.~Horn} \affiliation{South Dakota Science and Technology Authority, Sanford Underground Research Facility, Lead, SD 57754, USA} \affiliation{University of California Berkeley, Department of Physics, Berkeley, CA 94720, USA} 
\author{D.Q.~Huang} \affiliation{Brown University, Department of Physics, 182 Hope St., Providence, RI 02912, USA}  
\author{C.M.~Ignarra} \affiliation{SLAC National Accelerator Laboratory, 2575 Sand Hill Road, Menlo Park, CA 94205, USA} \affiliation{Kavli Institute for Particle Astrophysics and Cosmology, Stanford University, 452 Lomita Mall, Stanford, CA 94309, USA} 
\author{R.G.~Jacobsen} \affiliation{University of California Berkeley, Department of Physics, Berkeley, CA 94720, USA}  
\author{O.~Jahangir} \affiliation{Department of Physics and Astronomy, University College London, Gower Street, London WC1E 6BT, United Kingdom}  
\author{W.~Ji} \affiliation{SLAC National Accelerator Laboratory, 2575 Sand Hill Road, Menlo Park, CA 94205, USA} \affiliation{Kavli Institute for Particle Astrophysics and Cosmology, Stanford University, 452 Lomita Mall, Stanford, CA 94309, USA} 
\author{K.~Kamdin} \affiliation{University of California Berkeley, Department of Physics, Berkeley, CA 94720, USA} \affiliation{Lawrence Berkeley National Laboratory, 1 Cyclotron Rd., Berkeley, CA 94720, USA} 
\author{K.~Kazkaz} \affiliation{Lawrence Livermore National Laboratory, 7000 East Ave., Livermore, CA 94551, USA}  
\author{D.~Khaitan} \affiliation{University of Rochester, Department of Physics and Astronomy, Rochester, NY 14627, USA}  
\author{E.V.~Korolkova} \affiliation{University of Sheffield, Department of Physics and Astronomy, Sheffield, S3 7RH, United Kingdom}  
\author{S.~Kravitz} \affiliation{Lawrence Berkeley National Laboratory, 1 Cyclotron Rd., Berkeley, CA 94720, USA}  
\author{V.A.~Kudryavtsev} \affiliation{University of Sheffield, Department of Physics and Astronomy, Sheffield, S3 7RH, United Kingdom}  
\author{E.~Leason} \affiliation{SUPA, School of Physics and Astronomy, University of Edinburgh, Edinburgh EH9 3FD, United Kingdom}  
\author{B.G.~Lenardo} \affiliation{University of California Davis, Department of Physics, One Shields Ave., Davis, CA 95616, USA} \affiliation{Lawrence Livermore National Laboratory, 7000 East Ave., Livermore, CA 94551, USA} 
\author{K.T.~Lesko} \affiliation{Lawrence Berkeley National Laboratory, 1 Cyclotron Rd., Berkeley, CA 94720, USA}  
\author{J.~Liao} \affiliation{Brown University, Department of Physics, 182 Hope St., Providence, RI 02912, USA}  
\author{J.~Lin} \affiliation{University of California Berkeley, Department of Physics, Berkeley, CA 94720, USA}  
\author{A.~Lindote} \affiliation{LIP-Coimbra, Department of Physics, University of Coimbra, Rua Larga, 3004-516 Coimbra, Portugal}  
\author{M.I.~Lopes} \affiliation{LIP-Coimbra, Department of Physics, University of Coimbra, Rua Larga, 3004-516 Coimbra, Portugal}  
\author{A.~Manalaysay} \affiliation{University of California Davis, Department of Physics, One Shields Ave., Davis, CA 95616, USA}  
\author{R.L.~Mannino} \affiliation{Texas A \& M University, Department of Physics, College Station, TX 77843, USA} \affiliation{University of Wisconsin-Madison, Department of Physics, 1150 University Ave., Madison, WI 53706, USA} 
\author{N.~Marangou} \affiliation{Imperial College London, High Energy Physics, Blackett Laboratory, London SW7 2BZ, United Kingdom}  
\author{M.F.~Marzioni} \affiliation{SUPA, School of Physics and Astronomy, University of Edinburgh, Edinburgh EH9 3FD, United Kingdom}  
\author{D.N.~McKinsey} \affiliation{University of California Berkeley, Department of Physics, Berkeley, CA 94720, USA} \affiliation{Lawrence Berkeley National Laboratory, 1 Cyclotron Rd., Berkeley, CA 94720, USA} 
\author{D.-M.~Mei} \affiliation{University of South Dakota, Department of Physics, 414E Clark St., Vermillion, SD 57069, USA}  
\author{M.~Moongweluwan} \affiliation{University of Rochester, Department of Physics and Astronomy, Rochester, NY 14627, USA}  
\author{J.A.~Morad} \affiliation{University of California Davis, Department of Physics, One Shields Ave., Davis, CA 95616, USA}  
\author{A.St.J.~Murphy} \affiliation{SUPA, School of Physics and Astronomy, University of Edinburgh, Edinburgh EH9 3FD, United Kingdom}  
\author{A.~Naylor} \affiliation{University of Sheffield, Department of Physics and Astronomy, Sheffield, S3 7RH, United Kingdom}  
\author{C.~Nehrkorn} \affiliation{University of California Santa Barbara, Department of Physics, Santa Barbara, CA 93106, USA}  
\author{H.N.~Nelson} \affiliation{University of California Santa Barbara, Department of Physics, Santa Barbara, CA 93106, USA}  
\author{F.~Neves} \affiliation{LIP-Coimbra, Department of Physics, University of Coimbra, Rua Larga, 3004-516 Coimbra, Portugal}  
\author{A.~Nilima} \affiliation{SUPA, School of Physics and Astronomy, University of Edinburgh, Edinburgh EH9 3FD, United Kingdom}  
\author{K.C.~Oliver-Mallory} \affiliation{University of California Berkeley, Department of Physics, Berkeley, CA 94720, USA} \affiliation{Lawrence Berkeley National Laboratory, 1 Cyclotron Rd., Berkeley, CA 94720, USA} 
\author{K.J.~Palladino} \affiliation{University of Wisconsin-Madison, Department of Physics, 1150 University Ave., Madison, WI 53706, USA}  
\author{E.K.~Pease} \affiliation{University of California Berkeley, Department of Physics, Berkeley, CA 94720, USA} \affiliation{Lawrence Berkeley National Laboratory, 1 Cyclotron Rd., Berkeley, CA 94720, USA} 
\author{Q.~Riffard} \affiliation{University of California Berkeley, Department of Physics, Berkeley, CA 94720, USA} \affiliation{Lawrence Berkeley National Laboratory, 1 Cyclotron Rd., Berkeley, CA 94720, USA} 
\author{G.R.C.~Rischbieter} \affiliation{University at Albany, State University of New York, Department of Physics, 1400 Washington Ave., Albany, NY 12222, USA}  
\author{C.~Rhyne} \affiliation{Brown University, Department of Physics, 182 Hope St., Providence, RI 02912, USA}  
\author{P.~Rossiter} \affiliation{University of Sheffield, Department of Physics and Astronomy, Sheffield, S3 7RH, United Kingdom}  
\author{S.~Shaw} \affiliation{University of California Santa Barbara, Department of Physics, Santa Barbara, CA 93106, USA} \affiliation{Department of Physics and Astronomy, University College London, Gower Street, London WC1E 6BT, United Kingdom} 
\author{T.A.~Shutt} \affiliation{SLAC National Accelerator Laboratory, 2575 Sand Hill Road, Menlo Park, CA 94205, USA} \affiliation{Kavli Institute for Particle Astrophysics and Cosmology, Stanford University, 452 Lomita Mall, Stanford, CA 94309, USA} 
\author{C.~Silva} \affiliation{LIP-Coimbra, Department of Physics, University of Coimbra, Rua Larga, 3004-516 Coimbra, Portugal}  
\author{M.~Solmaz} \affiliation{University of California Santa Barbara, Department of Physics, Santa Barbara, CA 93106, USA}  
\author{V.N.~Solovov} \affiliation{LIP-Coimbra, Department of Physics, University of Coimbra, Rua Larga, 3004-516 Coimbra, Portugal}  
\author{P.~Sorensen} \affiliation{Lawrence Berkeley National Laboratory, 1 Cyclotron Rd., Berkeley, CA 94720, USA}  
\author{T.J.~Sumner} \affiliation{Imperial College London, High Energy Physics, Blackett Laboratory, London SW7 2BZ, United Kingdom}  
\author{M.~Szydagis} \affiliation{University at Albany, State University of New York, Department of Physics, 1400 Washington Ave., Albany, NY 12222, USA}  
\author{D.J.~Taylor} \affiliation{South Dakota Science and Technology Authority, Sanford Underground Research Facility, Lead, SD 57754, USA}  
\author{R.~Taylor} \affiliation{Imperial College London, High Energy Physics, Blackett Laboratory, London SW7 2BZ, United Kingdom}  
\author{W.C.~Taylor} \affiliation{Brown University, Department of Physics, 182 Hope St., Providence, RI 02912, USA}  
\author{B.P.~Tennyson} \affiliation{Yale University, Department of Physics, 217 Prospect St., New Haven, CT 06511, USA}  
\author{P.A.~Terman} \affiliation{Texas A \& M University, Department of Physics, College Station, TX 77843, USA}  
\author{D.R.~Tiedt} \affiliation{University of Maryland, Department of Physics, College Park, MD 20742, USA}  
\author{W.H.~To} \affiliation{California State University Stanislaus, Department of Physics, 1 University Circle, Turlock, CA 95382, USA}  
\author{M.~Tripathi} \affiliation{University of California Davis, Department of Physics, One Shields Ave., Davis, CA 95616, USA}  
\author{L.~Tvrznikova} \affiliation{University of California Berkeley, Department of Physics, Berkeley, CA 94720, USA} \affiliation{Lawrence Berkeley National Laboratory, 1 Cyclotron Rd., Berkeley, CA 94720, USA} \affiliation{Yale University, Department of Physics, 217 Prospect St., New Haven, CT 06511, USA}
\author{U.~Utku} \affiliation{Department of Physics and Astronomy, University College London, Gower Street, London WC1E 6BT, United Kingdom}  
\author{S.~Uvarov} \affiliation{University of California Davis, Department of Physics, One Shields Ave., Davis, CA 95616, USA}  
\author{A.~Vacheret} \affiliation{Imperial College London, High Energy Physics, Blackett Laboratory, London SW7 2BZ, United Kingdom}  
\author{V.~Velan} \affiliation{University of California Berkeley, Department of Physics, Berkeley, CA 94720, USA}  
\author{R.C.~Webb} \affiliation{Texas A \& M University, Department of Physics, College Station, TX 77843, USA}  
\author{J.T.~White} \affiliation{Texas A \& M University, Department of Physics, College Station, TX 77843, USA}  
\author{T.J.~Whitis} \affiliation{SLAC National Accelerator Laboratory, 2575 Sand Hill Road, Menlo Park, CA 94205, USA} \affiliation{Kavli Institute for Particle Astrophysics and Cosmology, Stanford University, 452 Lomita Mall, Stanford, CA 94309, USA} 
\author{M.S.~Witherell} \affiliation{Lawrence Berkeley National Laboratory, 1 Cyclotron Rd., Berkeley, CA 94720, USA}  
\author{F.L.H.~Wolfs} \affiliation{University of Rochester, Department of Physics and Astronomy, Rochester, NY 14627, USA}  
\author{D.~Woodward} \affiliation{Pennsylvania State University, Department of Physics, 104 Davey Lab, University Park, PA  16802-6300, USA}  
\author{J.~Xu} \affiliation{Lawrence Livermore National Laboratory, 7000 East Ave., Livermore, CA 94551, USA}  
\author{C.~Zhang} \affiliation{University of South Dakota, Department of Physics, 414E Clark St., Vermillion, SD 57069, USA}

\collaboration{LUX Collaboration}

\date{\today}
\vspace{13 mm}
\begin{abstract}
\noindent We present the results of a direct detection search for mirror dark matter interactions, using data collected from the Large Underground Xenon experiment during 2013, with an exposure of 95 live-days $\times$ 118 kg. Here, the calculations of the mirror electron scattering rate in liquid xenon take into account the shielding effects from mirror dark matter captured within the Earth. Annual and diurnal modulation of the dark matter flux and atomic shell effects in xenon are also accounted for. Having found no evidence for an electron recoil signal induced by mirror dark matter interactions we place an upper limit on the kinetic mixing parameter over a range of local mirror electron temperatures between 0.1 and 0.9 keV. This limit shows significant improvement over the previous experimental constraint from orthopositronium decays and significantly reduces the allowed parameter space for the model. We exclude mirror electron temperatures above 0.3 keV at a 90\% confidence level, for this model, and constrain the kinetic mixing below this temperature.
\end{abstract}

\maketitle


\textit{Introduction ---}
The Standard Model (SM) is a gauge field theory with $SU(3)_c \bigotimes SU(2) \bigotimes U(1)$ gauge symmetry. It successfully describes known particles and their non-gravitational interactions, but does not contain a suitable dark matter candidate. One possibility for accommodating dark matter particles is that they exist in a hidden sector --- a collection of particles and fields which do not interact via SM gauge boson forces, but do interact with SM particles gravitationally ~\cite{Feng}.  Mirror dark matter is a special case where the hidden sector is exactly isomorphic to the SM~\cite{Foot2014}, having the same gauge symmetry. Therefore, it contains mirror partners (denoted~$'$) of the SM particles with the same masses, lifetimes and self interactions. 
The full Lagrangian may be written as:
\begin{equation}
\begin{split}
\Lagr = & \Lagr_{SM}(e, u, d, \gamma, W, Z, ...) + \\ & \Lagr_{SM}(e', u', d', \gamma', W', Z', ...) + \Lagr_{mix} ,
\end{split}
\end{equation}
where $\Lagr_{SM}(e,...)$ and $\Lagr_{SM}(e',...)$ are the Langrangians for the SM and mirror sectors, respectively. The two sectors are related by a discrete $Z_2$ symmetry transformation, with the only allowed non-gravitational interactions given by:
\begin{equation}
\Lagr_{mix} = \frac{\varepsilon}{2} F^{\mu \nu} F'_{\mu \nu} + \lambda \phi^{\dagger} \phi \phi^{' \dagger} \phi^{'}.
\end{equation}
Here, the first term describes kinetic mixing of $U(1)_Y$ and mirror $U(1)_Y'$, with field strength tensors $F_{\mu \nu}, F_{\mu \nu}'$ and kinetic mixing strength $\varepsilon$~\cite{Foot1991}. The second term describes Higgs ($\phi$) -- mirror Higgs ($\phi'$) mixing, with strength determined by parameter $\lambda$. Kinetic mixing induces tiny ordinary electric charges, $\pm \varepsilon e$ for the mirror protons and electrons~\cite{Holdom}. This allows very weak electromagnetic interactions between mirror and SM particles. The kinetic mixing parameter, $\varepsilon$, determines the strength of most mirror -- SM particle couplings and is thus the target of experimental searches. The Higgs -- mirror Higgs portal can be probed at colliders, through Higgs production and decays, but does not give observable signals in direct detection experiments \cite{Foot2014}.

Within the mirror dark matter model kinetic mixing is constrained theoretically to lie in the range; $10^{-11} \leq \varepsilon \leq 4 \times 10^{-10}$ \cite{Foot2014}. In order for the mirror dark matter halo to be in equilibrium, heating from supernovae must balance energy loss from dissipative processes, giving the lower limit on $\varepsilon$~\cite{FootTemp}. But if $\varepsilon$ is too high cosmic structure formation would be too heavily damped, giving the upper limit~\cite{FootVagnozzi}.

\textit{LUX Experiment ---}
The Large Underground Xenon (LUX) experiment was a dual phase (liquid-gas) time projection chamber (TPC), containing a 250 kg active mass of liquid xenon. The main aim of LUX was to search for dark matter in the form of weakly interacting massive particles (WIMPs), placing limits on spin-independent WIMP-nucleon cross-sections for WIMP masses above 4 GeV~\cite{LUXfirst, LUXreanalysis}. Other studies include searches for spin-dependent WIMP-nucleon interactions \cite{LUXspindep}, electron recoil searches for solar axions and axionlike particles~\cite{LUXaxion} and sub GeV dark matter via the Bremsstrahlung and Migdal effects~\cite{LUXsubGeV}.

As described in Ref. \cite{LUXdet}, the LUX TPC was located in a low-radioactivity titanium cryostat, itself within a 6.1 m high, 7.6 m diameter water tank 1458 m underground at the Sanford Underground Research Facility, Lead, USA. Details of the detector calibration and performance are available in Ref.~\cite{LUXcomp}.
When a particle interacts in the liquid xenon, prompt scintillation photons (S1) and ionisation electrons are produced. The ionisation electrons are drifted upwards by a vertical electric field and extracted into the gas phase, where they produce an electroluminescence signal (S2). Photons from these signals are detected by two arrays of 61 photomultiplier tubes, above and below the active volume. The (x,y) position is obtained from the S2 light distribution in the top PMT array and the depth is found from the delay of the S2 relative to the S1 \cite{LUXPos}, allowing for fiducialisation of the active volume.

The data used in this analysis was collected between 24th April and 1st September 2013, giving 118 kg $\times$ 95 live days total exposure. Four detector observables are used --- $r, z, S1_c, S2_c$, where $S1_c$ and $S2_c$ refer to amplitudes corrected to equalize the detector response throughout the active volume.

\textit{Signal Model ---}
Mirror dark matter would exist as a multi-component plasma halo, assuming that the mirror electron temperature exceeds the binding energy of a mirror hydrogen atom and the cooling time exceeds the Hubble time \cite{FootPlasma}. This halo is predominantly composed of mirror electrons, $\rm e'$, and mirror helium nuclei, $\rm He'$. The $\rm He'$ mass fraction is higher (and $\rm H'$ lower) than for ordinary matter because freeze out happens earlier, due to a lower initial temperature in the mirror sector \cite{Foot2014}. Kinetic mixing allows electromagnetic interactions between mirror and SM particles, meaning that mirror electrons in the halo can scatter off Xe atomic electrons in the LUX detector. 

For a dark matter halo in hydrostatic equilibrium, the local mirror electron temperature is given by ~\cite{FootTemp}: 
\begin{equation} \label{Eq:temp}
T = \frac{\overline{m} v_{rot}^2}{2},
\end{equation}
where $\overline{m}$ is the average mass of halo particles and $v_{rot}$ is the galactic rotational velocity. Arguments from early universe cosmology in the mirror model give a mirror helium mass fraction of 90\%~\cite{FootMass} and assuming a completely ionized plasma $\overline{m} \approx 1.1$ GeV. Therefore, using $v_{rot} \approx 220$ kms$^{-1}$ and assuming the halo is in hydrostatic equilibrium, local mirror electron temperature $\sim 0.3$ keV is expected.

In such plasma dark matter models, it is important to consider capture of the dark matter by the Earth~\cite{FootShield}. Mirror dark matter is captured when it loses energy due to kinetic mixing interactions with normal matter. Once a significant amount has accumulated, further capture occurs due to mirror dark matter self interactions. Subsequently, mirror dark matter will thermalize with normal matter in the Earth to form an extended distribution, which can affect the incoming mirror dark matter via collisional shielding or deflection by a dark ionosphere. Interactions with the dark ionosphere are very difficult to model ~\cite{FootPlasma}, but the collisional shielding, due to mirror particle interactions identical to the standard model version, can be accounted for. Here we follow the formalism presented in Ref. \cite{FootShield, FootPlasma, FootXenon}, first validating the calculations for NaI (as given in \cite{FootShield}) then performing the calculations for Xe.

The electron -- mirror electron Coulomb scattering cross section for this process is given by \cite{FootPlasma}:
\begin{equation} \label{eq:diffxs}
\frac{d\sigma}{dE_R} = \frac{\lambda}{E_R^2v^2}, \quad \quad \quad \lambda = \frac{2\pi \varepsilon^2 \alpha^2}{m_e}.
\end{equation}
Here $E_R$ is electron recoil energy, $v$ velocity of the incoming mirror electron, $m_e$ electron mass, $\varepsilon$ the kinetic mixing parameter and $\alpha$ the fine structure constant. The scattering rate, calculated by multiplying with the integral of the velocity distribution of the incoming mirror dark matter and Taylor expanding around the yearly average, is given by \cite{FootShield}:
\begin{equation} \label{eq:diffrate}
\begin{split}
\frac{dR}{dE_R} = g_T N_T n_{e'}^0 \frac{\lambda}{v_c^0 E_R^2} [1 &+ A_v\rm cos \omega(t - t_0) \\
& + A_{\theta} (\theta - \bar{\theta})].
\end{split}
\end{equation}
Here $N_T$ is the number of target electrons, $n_{e'}^0$ the number density of mirror electrons arriving at the detector and $v_c^0$ describes the modified velocity distribution at the detector due to shielding. The effective number of free electrons, $g_T$, is the number of electrons per target atom with atomic binding energy ($E_b$) less than recoil energy ($E_R$) --- modelled as a step function for the atomic shells in xenon. 

The $A_v\rm cos \omega(t - t_0)$ term describes annual modulation resulting from the change of velocity of the Earth with respect to the dark matter halo. Here $\omega = 2 \pi /$year, $t_0 = 153$ days (2nd June) and modulation amplitude $A_v=0.7$ \cite{FootShield}. The $A_{\theta} (\theta - \bar{\theta})$ term describes diurnal and annual modulation due to the rotation of the Earth and the variation of the Earth's spin axis relative to the incoming dark matter wind. Here $\theta$ is the angle between the halo wind and the zenith at the detector location, $\bar{\theta}$ is the yearly average and the amplitude is $A_{\theta}=1$. The time variation of $\theta$ is examined in \cite{FootPlasma}. The mean modulation terms over the data taking period, accounting for the live time per day, are $A_v\rm \langle cos \omega(t - t_0) \rangle = 0.556$ and $A_{\theta} \langle \theta - \bar{\theta} \rangle = 0.015$. 

Equation \ref{eq:diffxs} shows that $d\sigma/dE_R \propto 1/v^2$, so the collision length $\propto v^2$. This means that for sufficiently large incoming velocity, the effect of collisions becomes negligible (as scattering length exceeds the available distance). Therefore, above some cutoff velocity, $v_{cut}$, collisions do not need to be considered. Below this velocity collisions are important until mirror electron energy is reduced to $\sim$ 25 eV, after which energy loss to the captured mirror helium is no longer important. From energy loss considerations the cutoff velocity may be estimated as \cite{FootShield}:
\begin{equation}
v_{cut}^4 \approx \frac{16 \pi}{m_e^2} \alpha^2 \Sigma \rm log \Lambda,
\end{equation}
where $\Lambda \sim T/E_{min} \approx$ 20, with minimum collisional energy loss $E_{min}$. Column density, $\Sigma$, is calculated by integrating the number density of captured mirror helium nuclei over the path of the incoming mirror dark matter particle:
\begin{equation}
\Sigma(\psi) = \int n_{He'} \rm{dl}.
\end{equation}
Here $\psi$ is the angle between the direction of the incoming mirror electron and the zenith at the detectors location and $l$ is the distance travelled.

\begin{figure}[t!]
	\begin{center}
	\includegraphics[width=8.6cm]{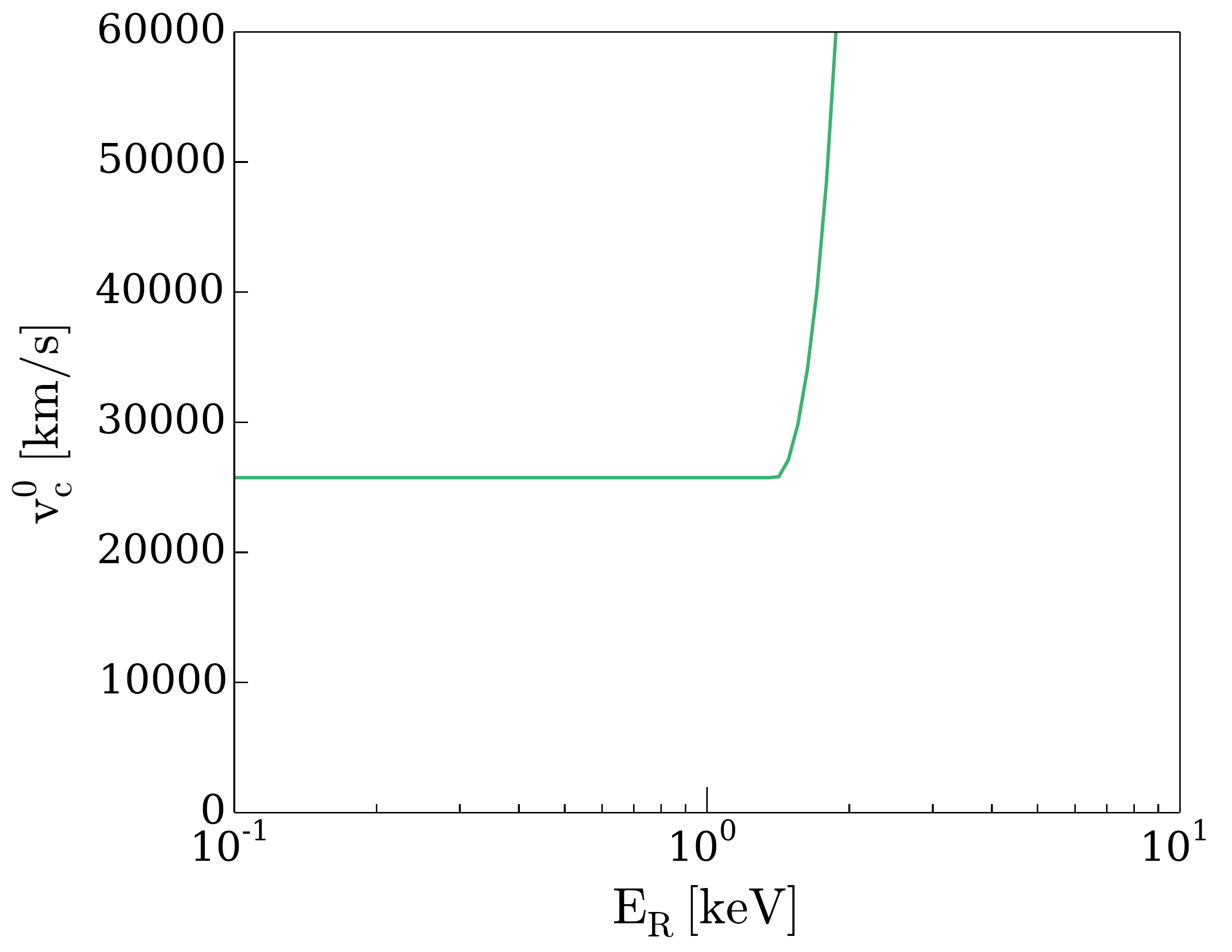} 
	\end{center}
	\caption{$v_c^0$ as a function of recoil energy; constant at low energy due to independence from $v_{min}$ rising steeply at higher energy where $v_{min}$ exceeds the mean particle velocity.}
	\label{fig:vc0}
\end{figure}

The energy dependent term describing the velocity distribution is given by \cite{FootShield}:
\begin{equation} \label{eq:vc0}
\frac{1}{v_c^0}  = \frac{1}{N v_0 \sqrt{\pi}} \int e^{-y^2/v_0^2} \rm{dcos\psi},
\end{equation}
where $v_0=\sqrt{2T/m_e}$ is the velocity dispersion. Dependence on recoil energy is through $y = MAX[v_{cut}(\psi), v_{min}(E_R)]$, where $v_{min}(E_R) = \sqrt{2E_R/m_e}$ is the minimum velocity needed to produce a recoil of energy $E_R$.

The dependence of $v_c^0$ on recoil energy is shown in Fig.~\ref{fig:vc0}. At low values of $E_R$ the average velocity exceeds the minimum, $|v | \gg v_{min}$, so most particles can produce recoils with energy $E_R$ and the integral becomes independent of $v_{min}$. For large $E_R$ the average particle velocity is lower than $v_{min}$, so the integral is suppressed, leading to a sharp rise in $v_c^0$. 

The normalization, $N$, is given by:
\begin{equation}
N = \int_{|v|>v_{cut}}^{\infty} \frac{e^{-v^2 /v_0^2}}{v_0^3 \pi^{3/2}} d^3v.
\end{equation}
The number density of the high velocity component which arrives at the Earth is given by:
\begin{equation} \label{eq:ne0}
n_{e'}^0 = N n_{e'}^{far},
\end{equation}
where $n_{e'}^{far}$ = 0.2 cm$^{-3}$ is the number density far from the Earth ~\cite{FootXenon}.

Both $v_c^0$ and $n_{e'}^0$ depend on the mirror helium density at the Earth's surface, $n_{He'}(R_E)$ (through column density), which is set to $n_{He'} = 5.8 \times 10^{-11} \rm {cm}^{-3}$  \cite{FootShield}. There is also dependence on electron recoil energy, $E_R$ (through $v_{min}$) and mirror electron temperature, $T$ (through $v_0$). Substituting Eq.~\ref{eq:vc0} and Eq.~\ref{eq:ne0} into Eq.~\ref{eq:diffrate} to calculate differential rate introduces dependence on the kinetic mixing parameter, $\varepsilon$ (through $\lambda$) and the target material (through $N_T$ and $g_T$). Calculation of the target independent parts $v_c^0$ and $n_{e'}^0$ was validated by evaluating the differential rate for NaI. This was convolved with the expected detector resolution, assumed to be Gaussian with energy dependent width \cite{DAMAExpt}, in order to reproduce Fig.4(a) from Ref.\cite{FootShield}.

\begin{figure}[t!]
	\begin{center}
	\includegraphics[width=8.6cm]{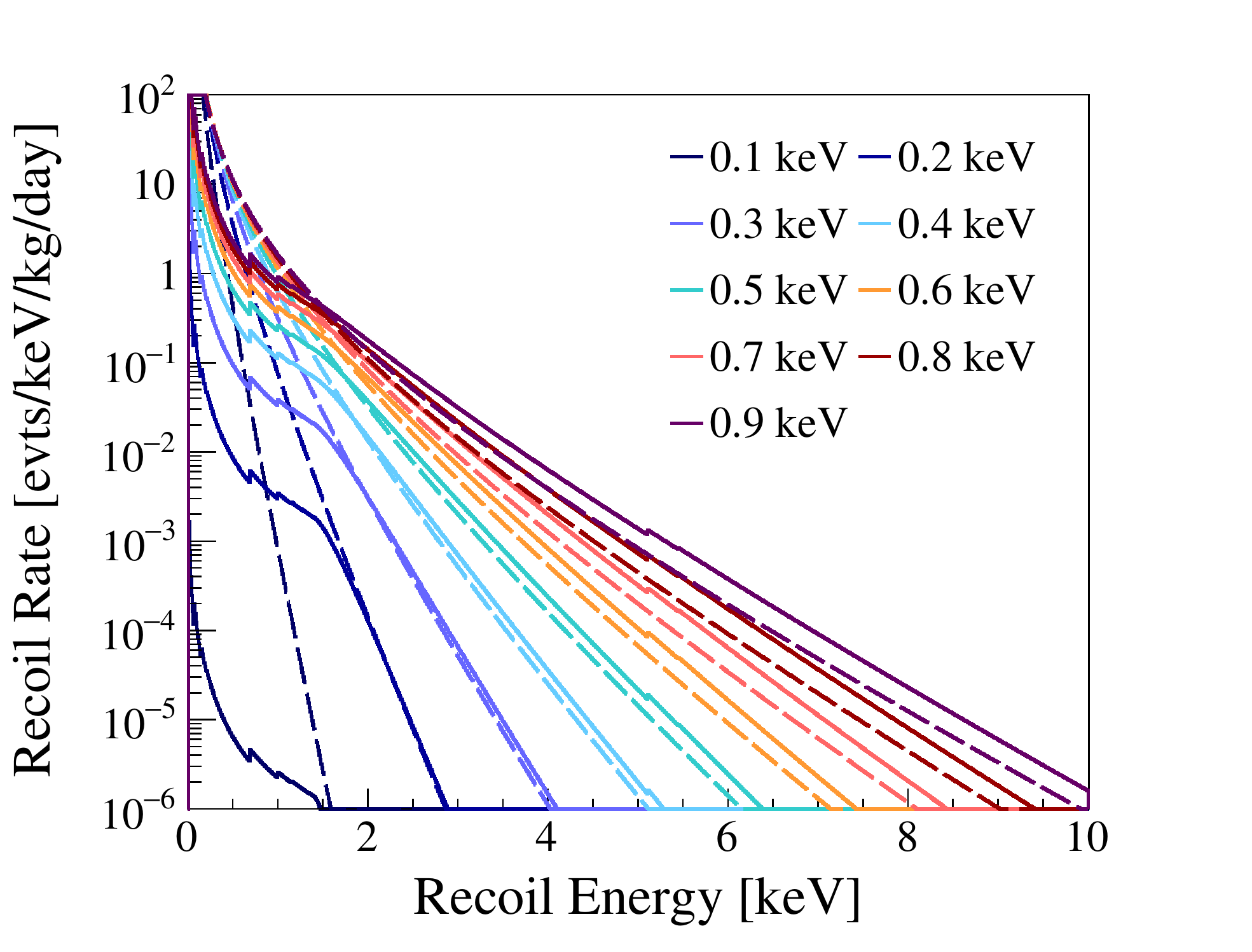} 
	\end{center}
	\caption{Electron recoil energy spectrum showing the differential rate of mirror electron scattering from xenon atomic electrons, with $\varepsilon = 10^{-10}$, both taking into account shielding effects (solid line) and with no shielding effects (dashed line).}
	\label{fig:spec}
\end{figure}

The differential rate of electron recoils in xenon could then be calculated using Eq.~\ref{eq:diffrate}. If the shielding effects are not accounted for a Maxwellian velocity distribution is assumed for the mirror electrons, with the rate given by Eq.(6.4) of Ref.~\cite{FootPlasma}. The differential energy spectra of electron recoils, calculated both with and without the shielding effects, are shown in Fig.~\ref{fig:spec} for a range of local mirror electron temperatures. 

\begin{figure}[t!]
	\centering
	\begin{subfigure}[b]{8.55cm}
	\includegraphics[width=8.55cm]{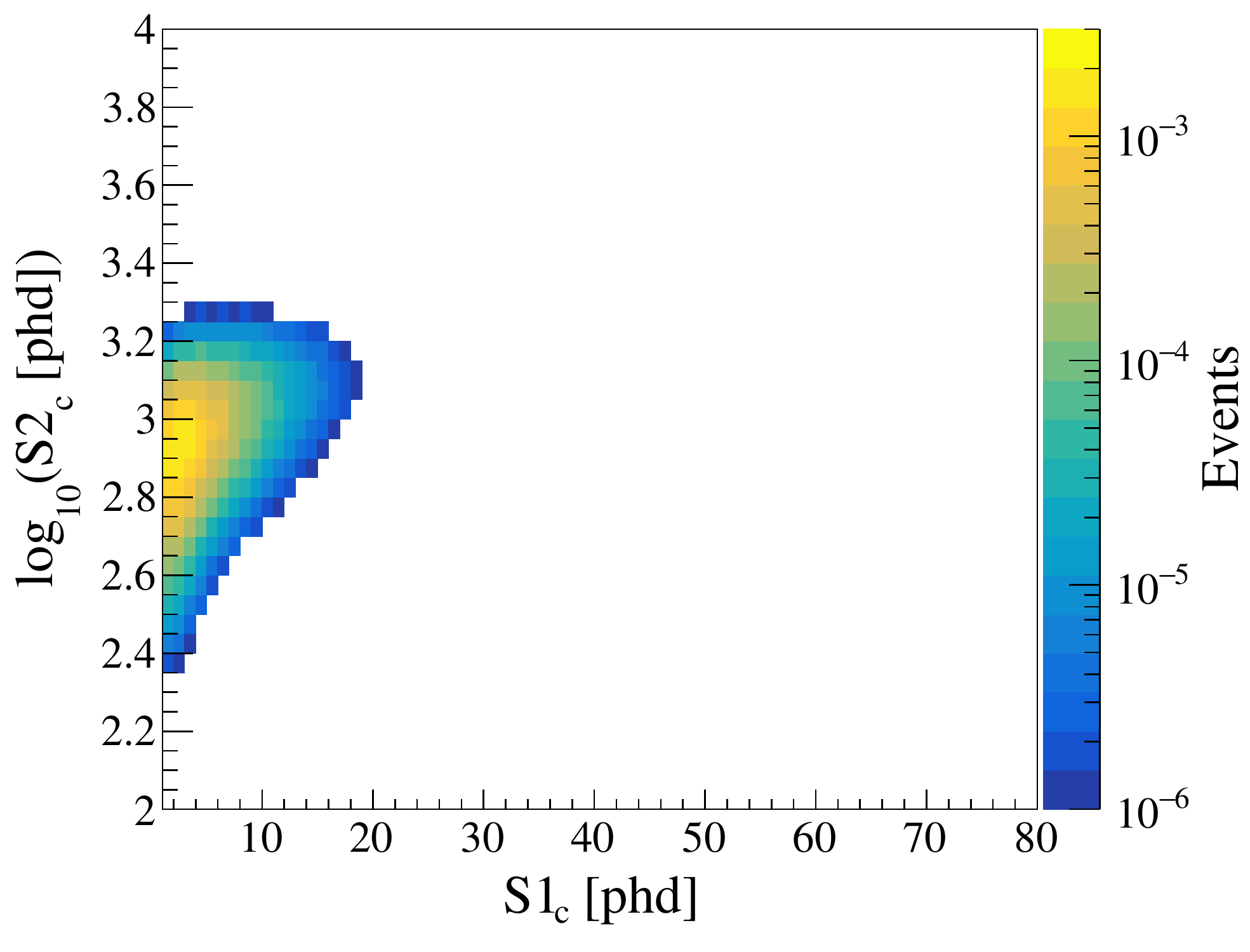}
	\caption{Signal model (T = 0.3 keV, $\varepsilon = 1 \times 10^{-10}$).}
	\label{fig:sig}
	\end{subfigure}
	\begin{subfigure}[b]{8.5cm}
	\includegraphics[width=8.5cm]{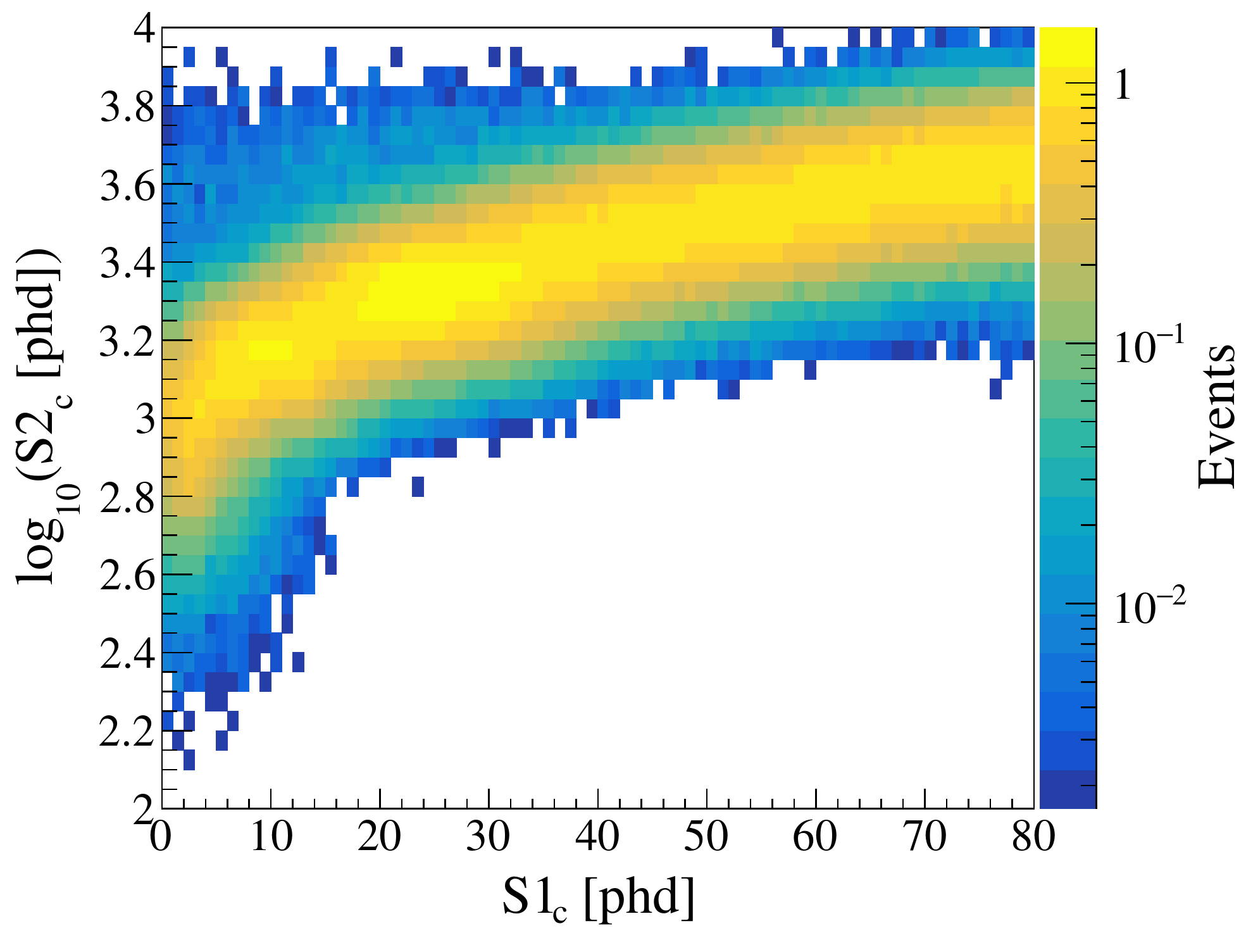}
	\caption{Background model}
	\label{fig:bkg}
	\end{subfigure}
	\caption{Signal and background model as projections of $\rm{log_{10}}$ $S2_c$ against S1.}
	\label{fig:sigbkg}
\end{figure}

The low energy electron recoil response of the LUX detector was characterised using an internal tritium calibration, as described in \cite{LUXTritium}. The injection of tritiated methane into the gas circulation gave a large sample of electron recoils from beta decays in the energy range of interest, used to precisely measure light and charge yields in the detector. These yields show good agreement with the Noble Element Simulation Technique (NEST) package v2.0 \cite{NESTv2}. Here we use NEST to model the distributions of the detector observables $r, z, S1_c, S2_c$, taking into account the detector resolution and efficiency, for signal events simulated using the above energy spectra. The quantities $S1_c$ and $S2_c$ are measured in photons detected (phd), with the resulting distribution in $\rm{log}_{10}$ $S2_c$ vs. $S1_c$ is shown in Fig.~\ref{fig:sig}, for mirror electron temperature $T=0.3$ keV and kinetic mixing $\varepsilon = 10^{-10}$.

\textit{Background Model ---}
Interactions of mirror dark matter particles within LUX would induce isolated low energy electron recoil events. Consequently, the signal being searched for competes with background events that arise from: Compton scattering of $\gamma$ rays from radioactive decay of isotopes in detector components, $\beta$ decay from $^{85m}$Kr and Rn contaminants in the liquid xenon and X-rays following $^{127}$Xe electron capture where the coincident $\gamma$ ray escapes detection~\cite{LUXradMu}.
Heavily down scattered decays from $^{238}$U chain, $^{232}$Th chain and $^{60}$Co generate additional $\gamma$ rays from the centre of a large copper block below the PMTs. The $\gamma$ rays can be modelled as two separate spatial distributions --- one from below the bottom PMT array and one from the rest of the detector. Decays of $^{37}$Ar, by electron capture, within the fiducial volume are also included \cite{LUXreanalysis}. A fiducial radius of 18 cm is used to exclude low energy events from $^{210}$Pb on the detector walls. The full background model used in this analysis is shown in Fig.~\ref{fig:bkg}, with each component normalized to the expected value.

\begin{figure}[t!]
	\begin{center}
	\includegraphics[width=8.6cm]{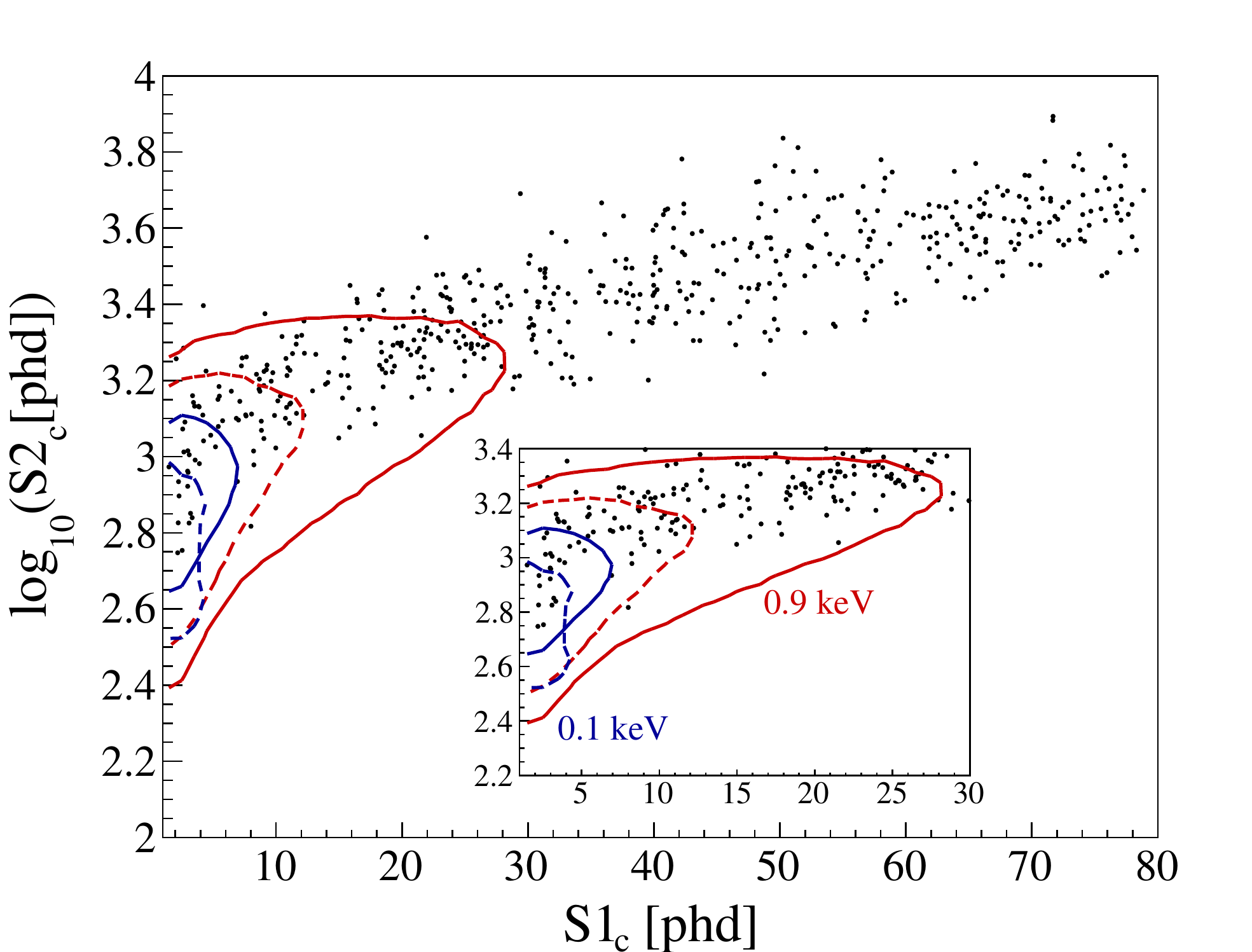}
	\end{center}
	\caption{LUX data with contours containing 90\% of the expected signal for mirror electron temperatures of 0.1 keV and 0.9 keV. Both are shown for kinetic mixing $\varepsilon = 10^{-10}$, the solid line with shielding effects and the dashed line without.}
	\label{fig:contour}
\end{figure}

\textit{Data Analysis ---}
A series of analysis cuts are applied to the data; events must also come from within a fiducial radius of 18 cm and z range of 8.5--48.6 cm above the bottom PMT array (drift time 305--38 $\mu$s). The S1 pulses in this analysis were required to have two PMTs in coincidence --- at least two non adjacent PMTs must measure an integrated area exceeding 0.3 phd. This is imposed to prevent spontaneous photocathode emission from being misidentified as an S1 pulse, as discussed in Ref. ~\cite{LUXcomp}. We also require $S1_c$ size 1--80 detected photons and the raw $S2$ size to exceed 165 detected photons. Corrected signal amplitudes $S1_c$, $S2_c$, account for non uniform temporal and spatial response throughout the detector, based on $^{83m}$Kr calibrations. Position corrections mean that it is possible to have an S1 size below 2 phd, despite this two fold coincidence requirement. The data cuts leave 516 events in our region of interested, shown in Fig.~\ref{fig:contour} along with 90\% signal contours. It should be noted that the signal model is not completely symmetric in $\rm{log_{10}}$ $S2_c$, so the contour containing 90\% of the signal will not be exactly centred on the ER band. This is a threshold effect due to the exponential shape of the signal model and is more pronounced for the sharply peaked signal models with no shielding.

The energy deposited by an event is given by \cite{LXeDetectors}:
\begin{equation}
    E = W (n_e + n_{\gamma}) = W \bigg( \frac{S1_c}{g_1} + \frac{S2_c}{g_2}\bigg),
\end{equation}
where $n_e$ and $n_{\gamma}$ are the number of electrons and photons produced, respectively and $W = (13.7 \pm 0.2)$ eV is the work function for producing these quanta in liquid xenon. Gain factors $g_1 = 0.117 \pm 0.003$ phd/photon and $g_2 = 12.1 \pm 0.8$ phd/electron were determined from calibrations \cite{LUXCalib}.

Compatibility with the data is tested using a two sided profile likelihood ratio test with four physics observables; $S1_c$, $\rm{log_{10}}$ $S2_c$, $r$, $z$ \cite{Cowan}. Simulated distributions of the signal model and background model were generated for each observable. The distribution of the test statistic, the ratio of the conditional maximum likelihood (with number of signal events fixed) to the global maximum likelihood, is found for a range of numbers of signal events. This is used to calculate the p-value for each number of signal events. The hypothesis test is then inverted to find the 90\% confidence limit on the number of signal events observed in the data. Systematic uncertainties in the background rates are treated as nuisance parameters. As detailed in Ref.~\cite{LUXradMu}, an extensive screening campaign gave the radioactive content of detector components, which was further constrained using data. Internal backgrounds were estimated from direct measurements of LUX data and sampling the Xe during the run. These were used to project the background rates for the period of data taking and normalize the Monte Carlo spectra. Nuisance parameters had the estimated rate as the mean value with a Gaussian constraint from the uncertainty. The best fit model covers zero signal model contribution for all mirror electron temperatures. The input and fit value for each nuisance parameter is shown in Table \ref{tab:PLRfit}, giving a total of $506 \pm 32$ background events, compared to 516 events in the data. For $T = 0.3$ keV, the background-only model gives KS test p-values of 0.27, 0.68, 0.71 and 0.60 for the projected distributions in $S1_c$, $\rm{log_{10}}$ $S2_c$, $r$ and $z$, respectively. For $T = 0.3$ keV this results in a 90\% confidence limit of 11 signal events, although it should be noted that the background events extend over a larger energy range than the signal.

\begin{table}[t!]
\centering
 \caption{Nuisance parameters used in the PLR test for a local mirror electron temperature 0.3 keV. The means and standard deviations of the Gaussian constraints are shown along with the value from the best fit to data.}
\begin{tabular}{l c c}
\hhline{===}
Parameter & \quad Constraint & \quad Fit Value  \\
\hline
\small{Low-z-origin $\gamma$ counts} & \quad 157 $\pm$ 78 \ & \quad 160 $\pm$ 17\\
\small{Other $\gamma$ counts} & \quad 217 $\pm$ 108 & \quad 179 $\pm$ 18  \\
\small{$\beta$ counts} & \quad 65 $\pm$ 32 & \quad 116 $\pm$ 17 \\
\small{$^{127}$Xe counts} & \quad 35 $\pm$ 18 & \quad 41 $\pm$ 8 \\
\small{$^{37}$Ar counts} & \quad 10 $\pm$ 5 & \quad 10 $\pm$ 7\\
\hline{}
\end{tabular}
\label{tab:PLRfit}
\end{table}

The 90\% confidence limit on kinetic mixing parameter is then calculated using:
\begin{equation}
\varepsilon (90 \% CL) = \varepsilon(0) \Bigg( \frac{nSig(90\% CL)}{nPDF(0)} \Bigg) ^{\frac{1}{2}},
\end{equation}
where $\varepsilon(0)$ is the arbitrary value of $\varepsilon$ used to generate the signal model, $nPDF(0)$ is the corresponding number of signal events and $nSig(90\% CL)$ is the 90\% confidence limit on the number of signal events. The power of $1/2$ comes from the dependence of the rate on $\varepsilon^2$ in Eq.~\ref{eq:diffxs}.

\begin{figure}[t!]
	\begin{center}
	\includegraphics[width=8.6cm]{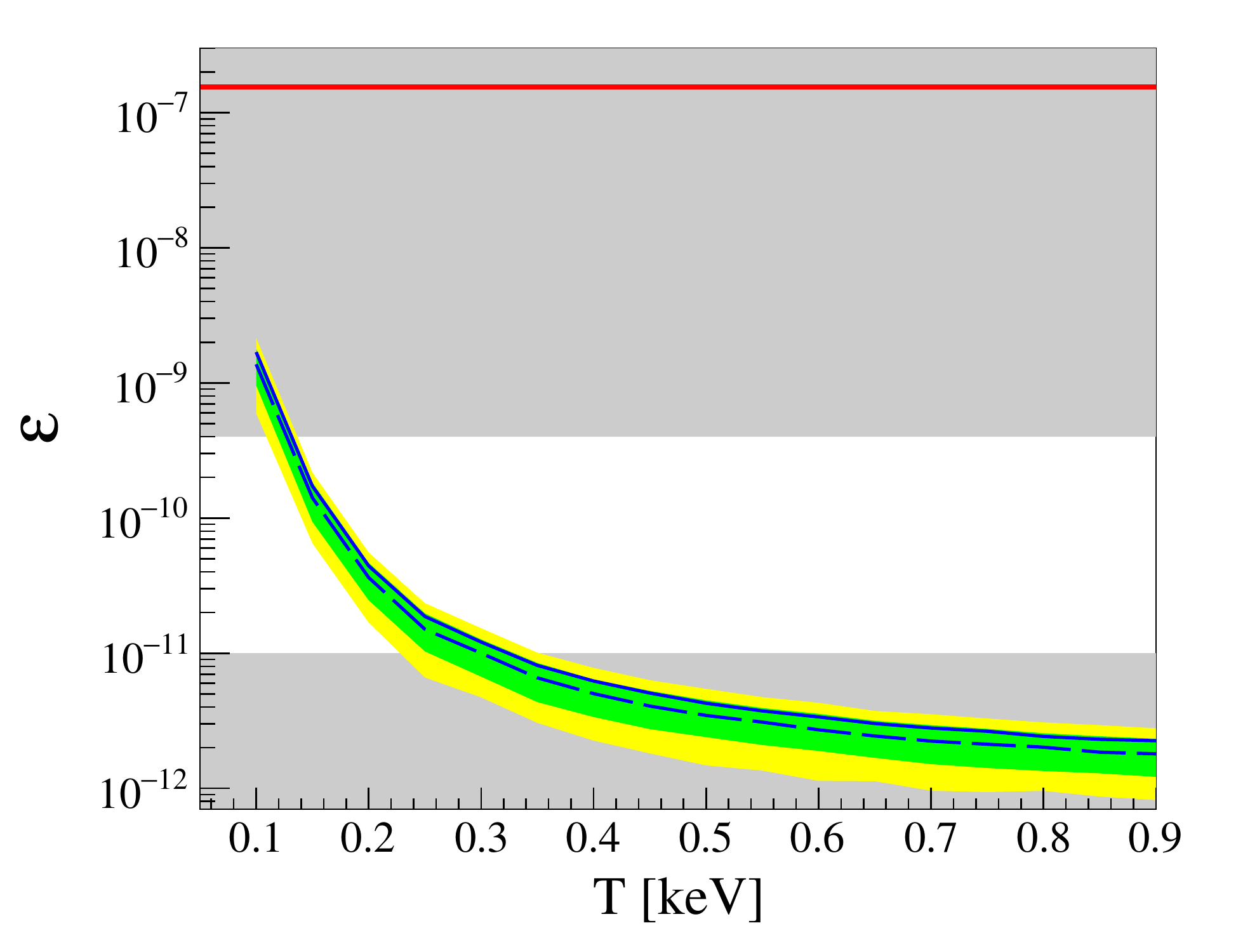}
	\end{center}
	\caption{Upper limit on kinetic mixing, at 90\% confidence level, as a function of local mirror electron temperature. The solid blue line shows this result, dashed blue is LUX sensitivity with green and yellow bands being 1 and 2 $\sigma$ respectively. The red line is the upper limit from orthopositronium decays \cite{oPs} and the grey regions are disallowed by the theory.}
	\label{fig:result}
\end{figure}

\textit{Results ---}
We set a 90\% confidence limit on the kinetic mixing parameter, $\varepsilon$, for the local mirror electron temperature range 0.1-0.9 keV, as shown in Fig.~\ref{fig:result}. The previous experimental constraint on $\varepsilon$ comes from invisible decays of orthopositronium in a vacuum~\cite{oPs}.
If positronium -- mirror positronium mixing were to occur, decay to missing photons would leave a missing energy signal. The upper limit placed on the branching fraction of orthopositronium to invisible states gives a 90\% upper confidence limit on the kinetic mixing parameter of $\varepsilon \leq 3.1 \times 10^{-7}$. The astrophysical constraint on kinetic mixing within the mirror dark matter theory; $10^{-11} \leq \varepsilon \leq 4 \times 10^{-10}$ \cite{Foot2014}, is also shown.

In Ref.~\cite{XENONleptophilic}, the XENON100 collaboration examine the possibility of leptophilic dark matter models explaining the DAMA \cite{DAMA2013} modulation signal. For each model the expected signal in xenon, given the DAMA modulation amplitude, is compared to XENON100 electron reocil data. This ruled out mirror dark matter as an explanation at a 3.6$\sigma$ confidence level,  but  there was no explicit search for mirror dark matter and no constraint was placed on the model itself.

\textit{Conclusion/Summary ---}
We have presented the results of the first dedicated direct detection search for mirror dark matter. The effect of mirror dark matter capture by the Earth and subsequent shielding is included, for the first time, for a signal in Xe. A significant proportion of the parameter space allowed by the theory is excluded by this analysis. However the present theoretical treatment makes assumptions for the local mirror electron temperature (thermal equilibrium with nuclei in the halo) and density \cite{FootPlasma,FootXenon}. The effect of deflection by the captured dark ionosphere is not included and this could significantly alter the signal model. Furthermore, the extent of these shielding effects may have significant dependence on the detector elevation relative to sea level, if the captured distribution is assumed to be spherically symmetric. 

Whilst there are possible caveats and extensions to this conceptually simple but phenomenologically complex mirror dark matter model, we have set limits based on the current model. This shows that it is possible to use direct detection experiments to probe low mass particles in a hidden sector.

\textit{Acknowledgements ---}
The authors would like to thank Robert Foot for helpful correspondence.

This Letter was partially supported by the U.S. Department of Energy (DOE) under Award No. DE-AC02-05CH11231, DE-AC05-06OR23100, DE-AC52-07NA27344, DE-FG01-91ER40618, DE-FG02-08ER41549, DE-FG02-11ER41738, DE-FG02-91ER40674, DE-FG02-91ER40688, DE-FG02-95ER40917, DE-NA0000979, DE-SC0006605, DE-SC0010010, DE-SC0015535, and DE-SC0019066; the U.S. National Science Foundation under Grants No. PHY-0750671, PHY-0801536, PHY-1003660, PHY-1004661, PHY-1102470, PHY-1312561, PHY-1347449, PHY-1505868, and PHY-1636738; the Research Corporation Grant No. RA0350; the Center for Ultra-low Background Experiments in the Dakotas (CUBED); and the South Dakota School of Mines and Technology (SDSMT). Laborat\'{o}rio de Instrumenta\c{c}\~{a}o e F\'{i}sica Experimental de Part\'{i}culas (LIP)-Coimbra acknowledges funding from Funda\c{c}\~{a}o para a Ci\^{e}ncia e a Tecnologia (FCT) through the Project-Grant PTDC/FIS-NUC/1525/2014. Imperial College and Brown University thank the UK Royal Society for travel funds under the International Exchange Scheme (IE120804). The UK groups acknowledge institutional support from Imperial College London, University College London and Edinburgh University, and from the Science \& Technology Facilities Council for for PhD studentships R504737 (EL), M126369B (NM), P006795 (AN), T93036D (RT) and N50449X (UU). This work was partially enabled by the University College London (UCL) Cosmoparticle Initiative. The University of Edinburgh is a charitable body, registered in Scotland, with Registration No. SC005336.\\
This research was conducted using computational resources and services at the Center for Computation and Visualization, Brown University, and also the Yale Science Research Software Core.\\
We gratefully acknowledge the logistical and technical support and the access to laboratory infrastructure provided to us by SURF and its personnel at Lead, South Dakota. SURF was developed by the South Dakota Science and Technology Authority, with an important philanthropic donation from T. Denny Sanford. Its operation is funded through Fermi National Accelerator Laboratory by the Department of Energy, Office of High Energy Physics.

\bibliographystyle{apsrev4-1}
\bibliography{mdm}

\begin{thebibliography}{28}%
\makeatletter
\providecommand \@ifxundefined [1]{%
 \@ifx{#1\undefined}
}%
\providecommand \@ifnum [1]{%
 \ifnum #1\expandafter \@firstoftwo
 \else \expandafter \@secondoftwo
 \fi
}%
\providecommand \@ifx [1]{%
 \ifx #1\expandafter \@firstoftwo
 \else \expandafter \@secondoftwo
 \fi
}%
\providecommand \natexlab [1]{#1}%
\providecommand \enquote  [1]{``#1''}%
\providecommand \bibnamefont  [1]{#1}%
\providecommand \bibfnamefont [1]{#1}%
\providecommand \citenamefont [1]{#1}%
\providecommand \href@noop [0]{\@secondoftwo}%
\providecommand \href [0]{\begingroup \@sanitize@url \@href}%
\providecommand \@href[1]{\@@startlink{#1}\@@href}%
\providecommand \@@href[1]{\endgroup#1\@@endlink}%
\providecommand \@sanitize@url [0]{\catcode `\\12\catcode `\$12\catcode
  `\&12\catcode `\#12\catcode `\^12\catcode `\_12\catcode `\%12\relax}%
\providecommand \@@startlink[1]{}%
\providecommand \@@endlink[0]{}%
\providecommand \url  [0]{\begingroup\@sanitize@url \@url }%
\providecommand \@url [1]{\endgroup\@href {#1}{\urlprefix }}%
\providecommand \urlprefix  [0]{URL }%
\providecommand \Eprint [0]{\href }%
\providecommand \doibase [0]{http://dx.doi.org/}%
\providecommand \selectlanguage [0]{\@gobble}%
\providecommand \bibinfo  [0]{\@secondoftwo}%
\providecommand \bibfield  [0]{\@secondoftwo}%
\providecommand \translation [1]{[#1]}%
\providecommand \BibitemOpen [0]{}%
\providecommand \bibitemStop [0]{}%
\providecommand \bibitemNoStop [0]{.\EOS\space}%
\providecommand \EOS [0]{\spacefactor3000\relax}%
\providecommand \BibitemShut  [1]{\csname bibitem#1\endcsname}%
\let\auto@bib@innerbib\@empty
\bibitem [{\citenamefont {Feng}\ \emph {et~al.}(2008)\citenamefont {Feng},
  \citenamefont {Tu},\ and\ \citenamefont {Yu}}]{Feng}%
  \BibitemOpen
  \bibfield  {author} {\bibinfo {author} {\bibfnamefont {J.}~\bibnamefont
  {Feng}}, \bibinfo {author} {\bibfnamefont {H.}~\bibnamefont {Tu}}, \ and\
  \bibinfo {author} {\bibfnamefont {H.}~\bibnamefont {Yu}},\ }\href {\doibase
  10.1088/1475-7516/2008/10/043} {\bibfield  {journal} {\bibinfo  {journal} {J.
  Cosmol. and Astropart. Phys.}\ }\textbf {\bibinfo {volume} {2008}},\ \bibinfo
  {pages} {1} (\bibinfo {year} {2008})},\ \Eprint
  {http://arxiv.org/abs/0808.2318} {arXiv:0808.2318} \BibitemShut {NoStop}%
\bibitem [{\citenamefont {Foot}(2014)}]{Foot2014}%
  \BibitemOpen
  \bibfield  {author} {\bibinfo {author} {\bibfnamefont {R.}~\bibnamefont
  {Foot}},\ }\href {\doibase 10.1142/S0217751X14300130} {\bibfield  {journal}
  {\bibinfo  {journal} {Int. J. Mod. Phys. A}\ }\textbf {\bibinfo {volume}
  {29}} (\bibinfo {year} {2014}),\ 10.1142/S0217751X14300130},\ \Eprint
  {http://arxiv.org/abs/1401.3965} {arXiv:1401.3965} \BibitemShut {NoStop}%
\bibitem [{\citenamefont {Foot}\ \emph {et~al.}(1991)\citenamefont {Foot},
  \citenamefont {Lew},\ and\ \citenamefont {Volkas}}]{Foot1991}%
  \BibitemOpen
  \bibfield  {author} {\bibinfo {author} {\bibfnamefont {R.}~\bibnamefont
  {Foot}}, \bibinfo {author} {\bibfnamefont {H.}~\bibnamefont {Lew}}, \ and\
  \bibinfo {author} {\bibfnamefont {R.}~\bibnamefont {Volkas}},\ }\href
  {\doibase 10.1016/0370-2693(91)91013-L} {\bibfield  {journal} {\bibinfo
  {journal} {Phys. Lett. B}\ }\textbf {\bibinfo {volume} {272}},\ \bibinfo
  {pages} {67} (\bibinfo {year} {1991})}\BibitemShut {NoStop}%
\bibitem [{\citenamefont {Holdom}(1986)}]{Holdom}%
  \BibitemOpen
  \bibfield  {author} {\bibinfo {author} {\bibfnamefont {B.}~\bibnamefont
  {Holdom}},\ }\href {\doibase 10.1016/0370-2693(86)91377-8} {\bibfield
  {journal} {\bibinfo  {journal} {Phys. Lett. B}\ }\textbf {\bibinfo {volume}
  {166}},\ \bibinfo {pages} {196} (\bibinfo {year} {1986})}\BibitemShut
  {NoStop}%
\bibitem [{\citenamefont {Foot}\ and\ \citenamefont {Volkas}(2004)}]{FootTemp}%
  \BibitemOpen
  \bibfield  {author} {\bibinfo {author} {\bibfnamefont {R.}~\bibnamefont
  {Foot}}\ and\ \bibinfo {author} {\bibfnamefont {R.}~\bibnamefont {Volkas}},\
  }\href {\doibase 10.1103/PhysRevD.70.123508} {\bibfield  {journal} {\bibinfo
  {journal} {Phys. Rev. D}\ }\textbf {\bibinfo {volume} {70}},\ \bibinfo
  {pages} {6} (\bibinfo {year} {2004})},\ \Eprint
  {http://arxiv.org/abs/astro-ph/0407522} {arXiv:astro-ph/0407522} \BibitemShut
  {NoStop}%
\bibitem [{\citenamefont {Foot}\ and\ \citenamefont
  {Vagnozzi}(2016)}]{FootVagnozzi}%
  \BibitemOpen
  \bibfield  {author} {\bibinfo {author} {\bibfnamefont {R.}~\bibnamefont
  {Foot}}\ and\ \bibinfo {author} {\bibfnamefont {S.}~\bibnamefont
  {Vagnozzi}},\ }\href {\doibase 10.1088/1475-7516/2016/07/013} {\bibfield
  {journal} {\bibinfo  {journal} {J. Cosmol. and Astropart. Phys.}\ }\textbf
  {\bibinfo {volume} {2016}} (\bibinfo {year} {2016}),\
  10.1088/1475-7516/2016/07/013},\ \Eprint {http://arxiv.org/abs/1602.02467}
  {arXiv:1602.02467} \BibitemShut {NoStop}%
\bibitem [{\citenamefont {Akerib}\ \emph {et~al.}(2014)\citenamefont {Akerib}
  \emph {et~al.}}]{LUXfirst}%
  \BibitemOpen
  \bibfield  {author} {\bibinfo {author} {\bibfnamefont {D.~S.}\ \bibnamefont
  {Akerib}} \emph {et~al.} (\bibinfo {collaboration} {LUX}),\ }\href {\doibase
  10.1103/PhysRevLett.112.091303} {\bibfield  {journal} {\bibinfo  {journal}
  {Phys. Rev. Lett.}\ }\textbf {\bibinfo {volume} {112}} (\bibinfo {year}
  {2014}),\ 10.1103/PhysRevLett.112.091303},\ \Eprint
  {http://arxiv.org/abs/1310.8214} {arXiv:1310.8214} \BibitemShut {NoStop}%
\bibitem [{\citenamefont {Akerib}\ \emph
  {et~al.}(2016{\natexlab{a}})\citenamefont {Akerib} \emph
  {et~al.}}]{LUXreanalysis}%
  \BibitemOpen
  \bibfield  {author} {\bibinfo {author} {\bibfnamefont {D.~S.}\ \bibnamefont
  {Akerib}} \emph {et~al.} (\bibinfo {collaboration} {LUX}),\ }\href {\doibase
  10.1103/PhysRevLett.116.161301} {\bibfield  {journal} {\bibinfo  {journal}
  {Phys. Rev. Lett.}\ }\textbf {\bibinfo {volume} {116}},\ \bibinfo {pages} {1}
  (\bibinfo {year} {2016}{\natexlab{a}})},\ \Eprint
  {http://arxiv.org/abs/1512.03506} {arXiv:1512.03506} \BibitemShut {NoStop}%
\bibitem [{\citenamefont {Akerib}\ \emph
  {et~al.}(2017{\natexlab{a}})\citenamefont {Akerib} \emph
  {et~al.}}]{LUXspindep}%
  \BibitemOpen
  \bibfield  {author} {\bibinfo {author} {\bibfnamefont {D.~S.}\ \bibnamefont
  {Akerib}} \emph {et~al.} (\bibinfo {collaboration} {LUX}),\ }\href {\doibase
  10.1103/PhysRevLett.118.251302} {\bibfield  {journal} {\bibinfo  {journal}
  {Phys. Rev. Lett.}\ }\textbf {\bibinfo {volume} {118}},\ \bibinfo {pages}
  {251302} (\bibinfo {year} {2017}{\natexlab{a}})}\BibitemShut {NoStop}%
\bibitem [{\citenamefont {Akerib}\ \emph
  {et~al.}(2017{\natexlab{b}})\citenamefont {Akerib} \emph
  {et~al.}}]{LUXaxion}%
  \BibitemOpen
  \bibfield  {author} {\bibinfo {author} {\bibfnamefont {D.~S.}\ \bibnamefont
  {Akerib}} \emph {et~al.} (\bibinfo {collaboration} {LUX}),\ }\href {\doibase
  10.1103/PhysRevLett.118.261301} {\bibfield  {journal} {\bibinfo  {journal}
  {Phys. Rev. Lett.}\ }\textbf {\bibinfo {volume} {118}},\ \bibinfo {pages}
  {261301} (\bibinfo {year} {2017}{\natexlab{b}})}\BibitemShut {NoStop}%
\bibitem [{\citenamefont {Akerib}\ \emph {et~al.}(2019)\citenamefont {Akerib}
  \emph {et~al.}}]{LUXsubGeV}%
  \BibitemOpen
  \bibfield  {author} {\bibinfo {author} {\bibfnamefont {D.~S.}\ \bibnamefont
  {Akerib}} \emph {et~al.} (\bibinfo {collaboration} {LUX}),\ }\href {\doibase
  10.1103/PhysRevLett.122.131301} {\bibfield  {journal} {\bibinfo  {journal}
  {Phys. Rev. Lett.}\ }\textbf {\bibinfo {volume} {122}},\ \bibinfo {pages}
  {131301} (\bibinfo {year} {2019})},\ \Eprint
  {http://arxiv.org/abs/1811.11241} {arXiv:1811.11241} \BibitemShut {NoStop}%
\bibitem [{\citenamefont {Akerib}\ \emph {et~al.}(2013)\citenamefont {Akerib}
  \emph {et~al.}}]{LUXdet}%
  \BibitemOpen
  \bibfield  {author} {\bibinfo {author} {\bibfnamefont {D.~S.}\ \bibnamefont
  {Akerib}} \emph {et~al.} (\bibinfo {collaboration} {LUX}),\ }\href {\doibase
  10.1016/j.nima.2012.11.135} {\bibfield  {journal} {\bibinfo  {journal} {Nucl.
  Instrum. Methods Phys. Res. A}\ }\textbf {\bibinfo {volume} {704}},\ \bibinfo
  {pages} {111} (\bibinfo {year} {2013})},\ \Eprint
  {http://arxiv.org/abs/1211.3788} {arXiv:1211.3788} \BibitemShut {NoStop}%
\bibitem [{\citenamefont {Akerib}\ \emph
  {et~al.}(2018{\natexlab{a}})\citenamefont {Akerib} \emph {et~al.}}]{LUXcomp}%
  \BibitemOpen
  \bibfield  {author} {\bibinfo {author} {\bibfnamefont {D.~S.}\ \bibnamefont
  {Akerib}} \emph {et~al.} (\bibinfo {collaboration} {LUX}),\ }\href
  {http://arxiv.org/abs/1712.05696} {\bibfield  {journal} {\bibinfo  {journal}
  {Phys. Rev. D}\ }\textbf {\bibinfo {volume} {97}},\ \bibinfo {pages} {1}
  (\bibinfo {year} {2018}{\natexlab{a}})},\ \Eprint
  {http://arxiv.org/abs/1712.05696} {arXiv:1712.05696} \BibitemShut {NoStop}%
\bibitem [{\citenamefont {Akerib}\ \emph
  {et~al.}(2018{\natexlab{b}})\citenamefont {Akerib} \emph {et~al.}}]{LUXPos}%
  \BibitemOpen
  \bibfield  {author} {\bibinfo {author} {\bibfnamefont {D.~S.}\ \bibnamefont
  {Akerib}} \emph {et~al.} (\bibinfo {collaboration} {LUX}),\ }\href {\doibase
  10.1088/1748-0221/13/02/p02001} {\bibfield  {journal} {\bibinfo  {journal}
  {J. Instrum.}\ }\textbf {\bibinfo {volume} {13}},\ \bibinfo {pages} {P02001}
  (\bibinfo {year} {2018}{\natexlab{b}})}\BibitemShut {NoStop}%
\bibitem [{\citenamefont {Clarke}\ and\ \citenamefont
  {Foot}(2016)}]{FootPlasma}%
  \BibitemOpen
  \bibfield  {author} {\bibinfo {author} {\bibfnamefont {J.}~\bibnamefont
  {Clarke}}\ and\ \bibinfo {author} {\bibfnamefont {R.}~\bibnamefont {Foot}},\
  }\href {\doibase 10.1088/1475-7516/2016/01/029} {\bibfield  {journal}
  {\bibinfo  {journal} {J. Cosmol. and Astropart. Phys.}\ }\textbf {\bibinfo
  {volume} {2016}} (\bibinfo {year} {2016}),\ 10.1088/1475-7516/2016/01/029},\
  \Eprint {http://arxiv.org/abs/1512.06471v1} {arXiv:1512.06471v1} \BibitemShut
  {NoStop}%
\bibitem [{\citenamefont {Ciarcelluti}\ and\ \citenamefont
  {Foot}(2010)}]{FootMass}%
  \BibitemOpen
  \bibfield  {author} {\bibinfo {author} {\bibfnamefont {P.}~\bibnamefont
  {Ciarcelluti}}\ and\ \bibinfo {author} {\bibfnamefont {R.}~\bibnamefont
  {Foot}},\ }\href {\doibase 10.1016/j.physletb.2010.06.003} {\bibfield
  {journal} {\bibinfo  {journal} {Phys. Lett. B}\ }\textbf {\bibinfo {volume}
  {690}},\ \bibinfo {pages} {462} (\bibinfo {year} {2010})},\ \Eprint
  {http://arxiv.org/abs/0809.4438} {arXiv:0809.4438} \BibitemShut {NoStop}%
\bibitem [{\citenamefont {Foot}(2019)}]{FootShield}%
  \BibitemOpen
  \bibfield  {author} {\bibinfo {author} {\bibfnamefont {R.}~\bibnamefont
  {Foot}},\ }\href {\doibase 10.1016/j.physletb.2018.12.063} {\bibfield
  {journal} {\bibinfo  {journal} {Phys. Lett. B}\ }\textbf {\bibinfo {volume}
  {789}},\ \bibinfo {pages} {592} (\bibinfo {year} {2019})},\ \Eprint
  {http://arxiv.org/abs/1806.04293v2} {arXiv:1806.04293v2} \BibitemShut
  {NoStop}%
\bibitem [{\citenamefont {Clarke}\ and\ \citenamefont
  {Foot}(2017)}]{FootXenon}%
  \BibitemOpen
  \bibfield  {author} {\bibinfo {author} {\bibfnamefont {J.}~\bibnamefont
  {Clarke}}\ and\ \bibinfo {author} {\bibfnamefont {R.}~\bibnamefont {Foot}},\
  }\href {\doibase 10.1016/j.physletb.2016.12.047} {\bibfield  {journal}
  {\bibinfo  {journal} {Phys. Lett. B}\ }\textbf {\bibinfo {volume} {766}},\
  \bibinfo {pages} {29} (\bibinfo {year} {2017})},\ \Eprint
  {http://arxiv.org/abs/1606.09063v1} {arXiv:1606.09063v1} \BibitemShut
  {NoStop}%
\bibitem [{\citenamefont {Bernabei}\ \emph {et~al.}(2008)\citenamefont
  {Bernabei} \emph {et~al.}}]{DAMAExpt}%
  \BibitemOpen
  \bibfield  {author} {\bibinfo {author} {\bibfnamefont {R.}~\bibnamefont
  {Bernabei}} \emph {et~al.},\ }\href {\doibase
  https://doi.org/10.1016/j.nima.2008.04.082} {\bibfield  {journal} {\bibinfo
  {journal} {Nucl. Instrum. Methods Phys. Res.}\ }\textbf {\bibinfo {volume}
  {592}},\ \bibinfo {pages} {297 } (\bibinfo {year} {2008})}\BibitemShut
  {NoStop}%
\bibitem [{\citenamefont {Akerib}\ \emph
  {et~al.}(2016{\natexlab{b}})\citenamefont {Akerib} \emph
  {et~al.}}]{LUXTritium}%
  \BibitemOpen
  \bibfield  {author} {\bibinfo {author} {\bibfnamefont {D.~S.}\ \bibnamefont
  {Akerib}} \emph {et~al.} (\bibinfo {collaboration} {LUX}),\ }\href {\doibase
  10.1103/physrevd.93.072009} {\bibfield  {journal} {\bibinfo  {journal} {Phys.
  Rev. D}\ }\textbf {\bibinfo {volume} {93}},\ \bibinfo {pages} {1} (\bibinfo
  {year} {2016}{\natexlab{b}})}\BibitemShut {NoStop}%
\bibitem [{\citenamefont {Szydagis}\ \emph {et~al.}(2018)\citenamefont
  {Szydagis} \emph {et~al.}}]{NESTv2}%
  \BibitemOpen
  \bibfield  {author} {\bibinfo {author} {\bibfnamefont {M.}~\bibnamefont
  {Szydagis}} \emph {et~al.},\ }\href {\doibase 10.5281/zenodo.1314669} {\
  (\bibinfo {year} {2018}),\ 10.5281/zenodo.1314669}\BibitemShut {NoStop}%
\bibitem [{\citenamefont {Akerib}\ \emph {et~al.}(2015)\citenamefont {Akerib}
  \emph {et~al.}}]{LUXradMu}%
  \BibitemOpen
  \bibfield  {author} {\bibinfo {author} {\bibfnamefont {D.~S.}\ \bibnamefont
  {Akerib}} \emph {et~al.} (\bibinfo {collaboration} {LUX}),\ }\href {\doibase
  10.1016/j.astropartphys.2014.07.009} {\bibfield  {journal} {\bibinfo
  {journal} {Astropart. Phys.}\ }\textbf {\bibinfo {volume} {62}},\ \bibinfo
  {pages} {33} (\bibinfo {year} {2015})},\ \Eprint
  {http://arxiv.org/abs/1403.1299} {arXiv:1403.1299} \BibitemShut {NoStop}%
\bibitem [{\citenamefont {Aprile}\ and\ \citenamefont
  {Doke}(2010)}]{LXeDetectors}%
  \BibitemOpen
  \bibfield  {author} {\bibinfo {author} {\bibfnamefont {E.}~\bibnamefont
  {Aprile}}\ and\ \bibinfo {author} {\bibfnamefont {T.}~\bibnamefont {Doke}},\
  }\href {\doibase 10.1103/RevModPhys.82.2053} {\bibfield  {journal} {\bibinfo
  {journal} {Rev. Mod. Phys.}\ }\textbf {\bibinfo {volume} {82}},\ \bibinfo
  {pages} {2053} (\bibinfo {year} {2010})}\BibitemShut {NoStop}%
\bibitem [{\citenamefont {Akerib}\ \emph
  {et~al.}(2017{\natexlab{c}})\citenamefont {Akerib} \emph
  {et~al.}}]{LUXCalib}%
  \BibitemOpen
  \bibfield  {author} {\bibinfo {author} {\bibfnamefont {D.~S.}\ \bibnamefont
  {Akerib}} \emph {et~al.} (\bibinfo {collaboration} {LUX}),\ }\href {\doibase
  10.1103/PhysRevD.96.112011} {\bibfield  {journal} {\bibinfo  {journal} {Phys.
  Rev. D}\ }\textbf {\bibinfo {volume} {97}},\ \bibinfo {pages} {1} (\bibinfo
  {year} {2017}{\natexlab{c}})},\ \Eprint {http://arxiv.org/abs/1709.00800}
  {arXiv:1709.00800} \BibitemShut {NoStop}%
\bibitem [{\citenamefont {Cowan}\ \emph {et~al.}(2011)\citenamefont {Cowan},
  \citenamefont {Cranmer}, \citenamefont {Gross},\ and\ \citenamefont
  {Vitells}}]{Cowan}%
  \BibitemOpen
  \bibfield  {author} {\bibinfo {author} {\bibfnamefont {G.}~\bibnamefont
  {Cowan}}, \bibinfo {author} {\bibfnamefont {K.}~\bibnamefont {Cranmer}},
  \bibinfo {author} {\bibfnamefont {E.}~\bibnamefont {Gross}}, \ and\ \bibinfo
  {author} {\bibfnamefont {O.}~\bibnamefont {Vitells}},\ }\href {\doibase
  10.1140/epjc/s10052-011-1554-0} {\bibfield  {journal} {\bibinfo  {journal}
  {Eur. Phys. J. C}\ }\textbf {\bibinfo {volume} {71}} (\bibinfo {year}
  {2011}),\ 10.1140/epjc/s10052-011-1554-0},\ \Eprint
  {http://arxiv.org/abs/1007.1727} {arXiv:1007.1727} \BibitemShut {NoStop}%
\bibitem [{\citenamefont {Vigo}\ \emph {et~al.}(2018)\citenamefont {Vigo} \emph
  {et~al.}}]{oPs}%
  \BibitemOpen
  \bibfield  {author} {\bibinfo {author} {\bibfnamefont {C.}~\bibnamefont
  {Vigo}} \emph {et~al.},\ }\href {\doibase 10.1103/PhysRevD.97.092008}
  {\bibfield  {journal} {\bibinfo  {journal} {Phys. Rev. D}\ }\textbf {\bibinfo
  {volume} {97}},\ \bibinfo {pages} {092008} (\bibinfo {year} {2018})},\
  \Eprint {http://arxiv.org/abs/1803.05744} {arXiv:1803.05744} \BibitemShut
  {NoStop}%
\bibitem [{\citenamefont {Aprile}\ \emph {et~al.}(2015)\citenamefont {Aprile}
  \emph {et~al.}}]{XENONleptophilic}%
  \BibitemOpen
  \bibfield  {author} {\bibinfo {author} {\bibfnamefont {E.}~\bibnamefont
  {Aprile}} \emph {et~al.} (\bibinfo {collaboration} {Collaboration, The
  XENON}),\ }\href {\doibase 10.1126/science.aab2069} {\bibfield  {journal}
  {\bibinfo  {journal} {Science}\ }\textbf {\bibinfo {volume} {349}},\ \bibinfo
  {pages} {851} (\bibinfo {year} {2015})},\ \Eprint
  {http://arxiv.org/abs/https://science.sciencemag.org/content/349/6250/851.full.pdf}
  {https://science.sciencemag.org/content/349/6250/851.full.pdf} \BibitemShut
  {NoStop}%
\bibitem [{\citenamefont {Bernabei}\ \emph {et~al.}(2013)\citenamefont
  {Bernabei} \emph {et~al.}}]{DAMA2013}%
  \BibitemOpen
  \bibfield  {author} {\bibinfo {author} {\bibfnamefont {R.}~\bibnamefont
  {Bernabei}} \emph {et~al.},\ }\href {\doibase 10.1140/epjc/s10052-013-2648-7}
  {\bibfield  {journal} {\bibinfo  {journal} {The European Physical Journal C}\
  }\textbf {\bibinfo {volume} {73}},\ \bibinfo {pages} {2648} (\bibinfo {year}
  {2013})}\BibitemShut {NoStop}%
\end{thebibliography}%

\end{document}